# AN OBSERVATIONAL LIMIT ON CIRCUMSTELLAR Hα FROM SUPERNOVA 1994D

ROBERT J. CUMMING AND PETER LUNDQVIST
*Stockholm Observatory*
*S-133 36 Saltsjöbaden, Sweden*

LINDA J. SMITH
*Department of Physics & Astronomy*
*University College London, London WC1E 6BT, UK*

AND

MAX PETTINI AND DAVID L. KING
*Royal Greenwich Observatory*
*Madingley Road, Cambridge CB3 0EZ, UK*

**Abstract.** We searched for narrow Hα in a high-resolution spectrum of SN 1994D taken 10 days before maximum, and found none. We estimate the limit this places on the progenitor mass loss, and find that it is competitive with recent radio limits, and excludes the highest-mass-loss-rate symbiotic systems as possible progenitors of the normal Type Ia SN 1994D.

## 1. Introduction

Despite our understanding of the light curves and spectra of SN Ia, we still don't know the nature of their progenitor systems. Ideally, we would like to make a direct detection of progenitor system material. We describe here an attempt at such a detection.

Only a few SNe Ia have shown any evidence at all of a circumstellar medium (CSM), and in all cases serious doubt remains. Branch *et al.* (1983; also Wheeler 1992a) identified unresolved Hα emission in a spectrum of SN 1981B taken 6 days after $B$ maximum (+6 days). The line is, however, blueshifted by 2000 km s$^{-1}$ from the local interstellar Ca II absorption, making the identification hard to believe. Graham *et al.* (1983)

and Graham & Meikle (1986) reported the detection of an IR excess from SN 1982E. This SN was never classified, but exploded in an S0 galaxy, suggesting, at the time, that it was a Type Ia. Type Ib and II events have since been seen in S0 galaxies. Polcaro & Viotti (1991) reported narrow H$\alpha$ absorption from SN 1990M at $-4$d, though an earlier spectrum suggests that the absorption came from the galaxy rather than the SN (Della Valle, Benetti & Panagia 1996). In the case of SN 1991bg, no narrow lines were seen at $+1$d (Leibundgut et al. 1993), but narrow H$\alpha$ and Na I were identified at $+197$d by Ruiz-Lapuente et al. (1993). The narrow features may however just be Fe II lines (Branch et al. 1995; Turatto et al. 1996).

Branch et al. (1995) have reviewed the candidate progenitor systems for Type Ia supernovae. Where the white dwarf accretes primarily hydrogen from a red giant (or main sequence/subgiant) companion, by Roche lobe overflow or from a wind, the possibility exists of detecting signatures of circumstellar hydrogen from the SN. In order to reach the Chandrasekhar mass, the white dwarf needs to accrete at $\geq 10^{-7}$ M$_\odot$ yr$^{-1}$ (Branch et al. 1995). Since mass transfer from the companion to the white dwarf is probably only 10–30 percent efficient (Yungelson et al. 1995), the system as a whole loses mass at about $10^{-6}$ M$_\odot$ yr$^{-1}$. This material, if ionised as a result of the SN explosion, could be detectable as narrow H$\alpha$ emission or absorption (Wheeler 1992a,b). Alternatively, if the progenitor system is a supersoft X-ray source, lines might be detectable from a pre-existing ionised nebula (Rappaport et al. 1994; Branch et al. 1995).

In this paper, we report observational limits on circumstellar H$\alpha$ from the fairly normal Type Ia SN 1994D.

## 2. Observations and results

On 1994 March 10.14, we used the ISIS spectrograph with a CCD detector on the William Herschel Telescope to obtain a high-resolution ($R = 35$ km s$^{-1}$) spectrum of supernova 1994D, centred around H$\alpha$. The spectrum was briefly presented in King et al. (1995). The supernova was then only $6.5 \pm 1$ days old, and did not reach $B$ maximum until $10 \pm 1$ days later (Meikle et al. 1996). The spectrum is thus one of the earliest Type Ia spectra ever taken. Later high-resolution spectra of SN 1994D have been presented by Ho & Filippenko (1995), and by Patat et al. (1996).

We used a slit width of 1.2 arcsec passing though the SN and the centre of the host galaxy NGC 4526. The seeing, measured from the spatial profile of the supernova, was 1.4 arcsec. Two integrations of 1000s and 750s were added to produce the final spectrum. CCD reduction was carried out in the usual way, wavelength-calibrating with a CuAr arc lamp, and flux-calibrating by comparison with the flux standard Feige 34 (Stone 1977).

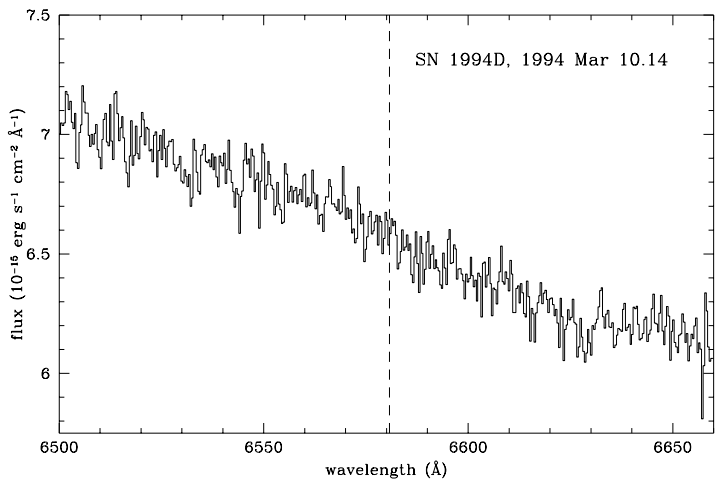

*Figure 1.*   Spectrum of SN 1994D around Hα. The broken line marks the expected velocity of the supernova, +830 km s$^{-1}$.

The absolute flux-calibration is probably accurate to around 10 percent.

The 2-D spectra show faint Hα emission extending out to only 0.8±0.3 arcsec from the supernova, with heliocentric velocity $v_{\rm HII}$ = +830 ± 50 km s$^{-1}$. This H II region emission provides our best estimate of the radial velocity of the SN itself. The velocities of local interstellar Na I absorption, +708 km s$^{-1}$, and the centre of the galaxy, +625 km s$^{-1}$ (King *et al.* 1995), are both lower. Figure 1 shows the high-resolution spectrum around Hα. The spectrum has not been dereddened: measurements of the Na I interstellar lines by King *et al.* (1995) and Ho & Filippenko (1995) give $E_{B-V}$ as 0.046 and $0.026^{+0.026}_{-0.013}$ respectively, corresponding to corrections in flux of only 5−10 percent, rather less than other errors in our results.

The two-dimensional data show that the H II region emission changes smoothly across the supernova position; there is no sign of any narrow emission or absorption from the supernova. We have therefore estimated 3-$\sigma$ detection limits for absorption and emission lines of various widths and velocities. We did this by adding Gaussian emission and absorption lines of various strengths, widths and velocities to the observed spectrum, and measuring the signal-to-noise ratio in the resulting line.

Detectable circumstellar material is likely to be at low velocity, so we expect CSM Hα to be unresolved in our spectra. For an unresolved line at the local H II region velocity, we found that the 3-$\sigma$ limit for emission flux is $2.0\times10^{-16}$ erg s$^{-1}$cm$^{-2}$, and the corresponding equivalent width limit for an absorption line is 0.028Å. If we instead allow the true SN velocity to differ by up to 200 km s$^{-1}$ from the local H II region velocity, these limits increase to $2.7 \times 10^{-16}$ erg s$^{-1}$cm$^{-2}$ and 0.037Å. The limits for resolved lines are correspondingly higher. For $v_{\rm FWHM} = 95$ km s$^{-1}$ the limits are instead $3.2 \times 10^{-16}$ erg s$^{-1}$cm$^{-2}$ and 0.062Å ($v_{\rm SN}=v_{\rm H\,II}$), or $5.4 \times 10^{-16}$ erg s$^{-1}$cm$^{-2}$ and 0.079Å ($-200$ km s$^{-1}$ < $v_{\rm SN} - v_{\rm H\,II}$ < $+200$ km s$^{-1}$). The data easily exclude an absorption line of the strength reported by Polcaro & Viotti (1991) from SN 1990M. Their line had $v_{\rm FWHM} = 760$ km s$^{-1}$ and an equivalent width of 1.3±0.2 Å; our 3-$\sigma$ limit for a line of this width from SN 1994D is 0.52Å.

## 3. Discussion

For us to see H$\alpha$ in emission, the CSM must be ionised. To see absorption, we require a large column density of neutral hydrogen excited to $n=2$ in front of the SN. How we interpret our limits therefore depends on the nature and strength of the ionising radiation. We consider three possible sources.

First, there is the burst of photons when the shock breaks out of the star. The number of ionising photons is, however, probably quite small, and the burst itself will be very short (around 1 second). For a black body with a radius of $2.9 \times 10^{-3}$ R$_\odot$ at a temperature of about $10^6$ K (T. Shigeyama, private communication), we expect only $\sim 3 \times 10^{47}$ photons with energies above 13.6 eV, enough only to ionise $\sim 2 \times 10^{-10}$ M$_\odot$ of hydrogen.

A day or so after the explosion, $\gamma$-rays from $^{56}$Ni decay in the SN ejecta emerge, but are also unlikely to ionise the CSM. Early on, the radiation is too hard, and the UV tail only shows itself after a few tens of days (P. Pinto, private communication), by which time the ejecta will have overrun the densest parts of the CSM.

The third possibility follows the suggestion of Chevalier (1984) that Type I supernovae whose progenitor white dwarfs have red giant or AGB star companions may interact with their surroundings in a similar manner to core-collapse SNe at late times (*e.g.* SNe 1979C and 1980K: Fransson 1982; Lundqvist & Fransson 1988; SN 1993J: Fransson, Lundqvist & Chevalier 1996). The supernova ejecta collide with the CSM, setting up a forward- and reverse-shock structure. Behind each shock the gas is heated to high temperatures and produces thermal X-ray emission. Being denser, the shocked SN gas radiates more strongly. For electron scattering optical depth $\gtrsim 10^{-3}$ in the shocked CSM, inverse Compton scattering of photospheric photons may also contribute to the ionising flux at very early times (*e.g.* Lundqvist & Fransson 1988). Nevertheless, the radiation from the shocked ejecta is the source of ionisation most likely to allow us to see narrow emission lines soon after explosion.

This third scenario is, we feel, the most promising way of producing cir-cumstellar *emission* lines from SN Ia. Assuming this is the case, we use our Hα limits to constrain the mass loss from SN 1994D's progenitor system. We adopt the simplest possible case, in which the supernova shock (velocity $\sim 3 \times 10^4$ km s$^{-1}$) heads out into a spherically symmetric red giant wind with an $r^{-2}$ density distribution. The X-rays from the shocked ejecta completely ionise the CSM. Assuming a temperature of $10^4$ K and case B recombination, and taking rates from Osterbrock (1989), we estimate the recombination emission luminosity in Hα for a given ratio of $\dot{M}/u$, where $u$ is the wind velocity. We take 16.2 Mpc as the distance to the SN, and 1994 March 3.5±1 as the time of explosion (Höflich 1995). The Hα flux fades approximately as $t^{-1}$. The limit on an unresolved emission line at $v_{\rm helio}=830$ km s$^{-1}$ gives $\dot{M}/u_{10} < 3 \times 10^{-6}$ (or $4 \times 10^{-6}$ if $-200$ km s$^{-1} < v_{\rm SN} - v_{\rm H\,II} < 200$ km s$^{-1}$), where $\dot{M}$ is the mass loss rate in M$_\odot$ yr$^{-1}$ and $u_{10}$ is the wind velocity in units of 10 km s$^{-1}$. If the temperature in the recombining gas is much higher than $10^4$ K, our limit becomes less useful, because the emission line will be thermally broadened. At $10^5$ K, for example, we expect $v_{\rm FWHM} \sim 70$ km s$^{-1}$, corresponding to a 3-$\sigma$ limit to $\dot{M}/u_{10}$ of $\sim 6 \times 10^{-6}$.

In principle, our absorption line limits can also be used to put limits on $\dot{M}/u$. The result, however, depends sensitively on the population in $n=2$ set up by the radiation first from the progenitor white dwarf and then from the SN itself. This requires careful modelling, and is beyond the scope of this paper.

$\dot{M}/u_{10}$ has been measured for about a hundred red giants in symbiotic systems (*e.g.* Seaquist & Taylor 1990; Mürset *et al.* 1991; Seaquist, Krogulec & Taylor 1992). The values lie in the range $10^{-8}$ to $2 \times 10^{-5}$; around two-thirds of the systems have $\dot{M}/u_{10}$ within a factor of 4 of $1.0 \times 10^{-7}$ (Seaquist *et al.* 1992). On the face of it, therefore, our limits for SN 1994D exclude only the 10 percent of symbiotic systems with the highest $\dot{M}/u$.

Limits on $\dot{M}/u$ (usually quoted as limits on $\dot{M}$) have been obtained for a few SN Ia by placing limits on radio and X-ray emission. The radio limit at 18 days for SN 1981B reported by Boffi & Branch (1995) suggested that $\dot{M}/u_{10}$ was probably less than about $10^{-6}$. Eck *et al.* (1995) reported that their radio non-detection of SN 1986G (distance 4.1 Mpc) was in 'clear conflict' with the range $10^{-7} \lesssim \dot{M}/u_{10} \lesssim 3 \times 10^{-6}$ at $-7$ days. Observing in X-rays, Schlegel & Petre (1993) established a limit for SN 1992A (16.9 Mpc) of $(2-3) \times 10^{-6}$ at $\sim 16$ days post-maximum, assuming that any X-rays would arise from circumstellar interaction. The limit on $\dot{M}/u$ from high-resolution optical spectroscopy scales roughly linearly with distance, so taking into account the difference in distance, our limit for SN 1994D is rather better than the radio limit for SN 1986G. However, since the radio, Hα and X-ray emission all probably decrease roughly as $t^{-1}$ after the explosion, these limits suggest that X-ray observations are the most promising way of actually detecting the CSM of a Type Ia supernova. At all wavebands, observation as soon as possible after discovery is crucial.

We find that upper limits on $\dot{M}/u$ based on a spherically symmetric Chevalier model are in general robust to the model's assumptions. The presence of its white dwarf companion means that mass loss from the red giant in a symbiotic system is unlikely to be spherically symmetric. Observations of those few systems which have circumstellar nebulae have generally shown bipolar structures on large and small scales (Solf & Ulrich 1985; Corradi & Schwarz 1993; Munari & Patat 1993). Aspherical mass loss necessarily affects the interpretation of limits on radio, X-ray and optical recombination emission. For example, taking optically thin emission, we estimate that if mass loss were equatorially concentrated into a solid angle $4\pi f$, where $f<1$, then assuming spherical symmetry in the Chevalier model will always *overestimate* the value of $\dot{M}/u$ by a factor of around $f^{-1/2}$, for at least radio and X-ray observations. The situation for optical lines is less clear.

A problem with radio emission in the Chevalier model is that the radio flux depends on the efficiency of converting kinetic energy of the SN blast wave to magnetic field energy and energy of relativistic electrons. This conversion is uncertain by a factor of about 10, corresponding to factor of 3−4 in $\dot{M}/u$. If synchrotron self-absorption is important for radio emission from SN Ia (as it seems to be for the early emission from SN 1987A; Chevalier & Dwarkadas 1995) then the uncertainty may be even greater.

Limits on early X-ray flux seem therefore likely to give the most reliable limits on $\dot{M}/u$ from the system, while $\dot{M}/u$ determined from radio limits will be uncertain, but to within a factor of only a few. Optical line limits are harder to interpret because of our ignorance of the CSM's ionisation state, but if detected, lines could provide much more powerful constraints on the composition, radiative acceleration, structure and geometry of the wind than either X-ray or radio observations.

Finally, our spectrum can in principle also test for another possible class of progenitor, the supersoft X-ray sources (SSXS; *e.g.* van den Heuvel *et al.* 1992). Some SSXS have surrounding photoionised nebulae, which may be detectable after explosion (Branch *et al.* 1995). We compared our limits at H$\alpha$ and [N II] $\lambda$6583 with predictions from Rappaport *et al.* (1994) and found that the maximum expected line flux in $\lambda$6583, the stronger of the two lines, is a factor of about three below our detection limit. The H$\alpha$ detected from the SSXS CAL 83 in the LMC (Remillard *et al.* 1995) would have been a factor of 30 below our limit. Future searches for such emission would be better carried out around [O III] $\lambda$5007, much the strongest optical line expected from SSXS nebulae.

**Acknowledgements** The data presented here are available as part of the La Palma archive of reduced supernova data (Martin *et al.* 1994; Meikle *et al.* 1995). We thank Eddie Baron,